%% file: main.tex
\documentclass{packages/quiin}

\usepackage{rotating}           
\usepackage[all,knot,arc,import,poly]{xy}   
\usepackage{soul}
\usepackage{svg}
\usepackage{threeparttable}
\usepackage{lscape}
\usepackage{lipsum}
\usepackage{cite}

\usepackage{longtable}
\usepackage{array}
\usepackage{longtable}

\usepackage{amsmath,amssymb,amsfonts}
\usepackage{graphicx}
\usepackage{textcomp}
\usepackage{xcolor}
\usepackage{url}
\usepackage{tikz}
\usepackage{listings}
\usetikzlibrary{arrows.meta,positioning,fit}
\idioma{EN} 

\tituloEN{Real-Time VPN Traffic over ETSI GS QKD 014 Key Delivery with a LuxQuanta NOVA QKD Platform}
\data{8}{06}{2026} 
\projeto{QuIIN Integração CV-QKD com Redes Clássicas}
\nomedoc{Experiment report}
\elaborado{Felipe Paixão <felipe.paixao@fbter.org.br> \\
    Anderson Altair Tomkelski <anderson.tomkelski@fieb.org.br> \\
    Marcus Elias Silva Freire <marcus.freire@fieb.org.br> \\
    Isys Nogueira de Sant'Anna <isys.nogueira@fbter.org.br> \\
    Adriano Humberto de Oliveira Maia <adriano.maia@fbter.org.br> \\
    Reinan da Silva Salazar <reinan.salazar@fieb.org.br> \\
    Ney Ricardo Lopez Junior <ney.junior@fbter.org.br> \\
    JOÃO Marcelo Silva Souza <joao.marcelo@fieb.org.br> \\
}
\revisado{}
\addrevisao{27/05/2026}{Escrita do relatório do experimento}
\addrevisao{08/06/2026}{Alteração do template do documento}

\loadglsentries[main]{tex/acronimos}

\incluilistadefiguras

\incluilistadetabelas

\incluilistadesiglas

\begin{document}

\textual

\input{tex/glossario}


\pagebreak

\chapter{Introduction}
Quantum Key Distribution (QKD) addresses a specific and important part of the secure-communications problem: the establishment of shared symmetric key material between distant sites. Unlike conventional public-key agreement, QKD does not base confidentiality of the generated key on computational hardness assumptions. Instead, it uses quantum states and authenticated classical post-processing to detect eavesdropping attempts and to distill common secret bits between two QKD endpoints \cite{qkd_kms_survey}. In deployed systems, those raw capabilities are not consumed directly by most applications. They are mediated by Key Management Entities (KMEs), which expose application-facing interfaces for obtaining key blocks and associated identifiers.

This separation between quantum key generation and classical application consumption is essential. A network encryptor, tunnel endpoint, or security management system normally requires an API that can be implemented, tested, logged, authenticated, and operated with conventional software practices. ETSI GS QKD 014 defines such an interface as a REST-based key delivery API using HTTPS and JSON \cite{etsi014}. It specifies how Secure Application Entities (SAEs) request fresh keys, how peer SAEs later retrieve matching keys by identifier, and which protocol elements are intentionally left to applications.

This experiment investigates this boundary from the perspective of a small but realistic consumer application: a Linux user-space VPN prototype. The VPN carries IP packets through a TUN interface, transports encrypted frames over UDP, and uses AES-256-GCM for per-packet authenticated encryption. Its distinguishing feature is that it does not rely on a pre-shared file of static keys. Instead, it obtains transmit keys from an ETSI GS QKD 014 KME and includes the returned \texttt{key\_ID} in every tunnel frame so that the receiving peer can retrieve the same key from its own KME.

The goal of the experiment is not to build a production VPN. Rather, it is to test whether a conventional network application can be made to follow the ETSI 014 master/slave key retrieval model with enough fidelity to interoperate first with a controlled simulator and later with real QKD infrastructure. The central contribution is therefore an implementation-oriented validation of the ETSI application pattern: key generation remains behind the KME boundary, while the VPN becomes responsible for transporting key identifiers, binding key identifiers to ciphertext, rotating keys, and rejecting malformed or replayed traffic.

\chapter{Background}
\section{QUANTUM KEY DISTRIBUTION AND CLASSICAL CONSUMERS}
QKD networks generate and distribute symmetric key material between trusted sites. A site typically contains QKD equipment, one or more KMEs, and one or more classical applications that consume keys. ETSI GS QKD 014 calls those consuming applications SAEs. Examples include optical encryption modules, security management systems, and network appliances \cite{etsi014}. This terminology is useful because it separates the application role from the physical QKD devices: the VPN in this experiment is an SAE, not a QKD transmitter or receiver.

The classical consumer still performs ordinary cryptographic operations. In this experiment, QKD-derived keys are used as AES-GCM keys. AES-GCM provides confidentiality and integrity for each encrypted frame, and it authenticates additional associated data (AAD) that is not itself encrypted \cite{nistgcm,rfc5116}. This is important because the VPN frame header contains routing and key-selection metadata. The payload is encrypted, while the header is authenticated as AAD so that tampering with fields such as \texttt{sender\_id}, sequence number, ciphertext length, or \texttt{key\_ID} causes decryption failure.

\section{REAL-WORLD QKD PROVIDERS}
The QKD industry contains several providers with field-oriented systems and integration tooling. ID Quantique offers production QKD systems such as Clavis XG and related key-management products, emphasizing long-distance fiber deployment, network topologies, and interoperability with encryptors \cite{idquantique-clavis}. Toshiba has developed QKD systems and QKD key-management products for quantum-secure networking, including multiplexed and long-distance variants intended for operation over optical infrastructure \cite{toshiba-qkd}. ThinkQuantum provides QKD systems for fiber, free-space, and hybrid channels, with products based on BB84-style QKD and QRNG-supported key generation \cite{thinkquantum-quky}. LuxQuanta, the provider used for the real backend stage of this experiment, develops Continuous-Variable QKD (CV-QKD) systems. Its NOVA LQ platform is presented as a deployable QKD system designed to coexist with classical optical communication over existing fiber infrastructure \cite{luxquanta-nova}.

These providers illustrate the practical motivation for using a standardized application API. Hardware and optical-layer implementations differ substantially, but a VPN, encryptor, or security appliance should not need to embed vendor-specific quantum protocol logic. It should instead consume authenticated key containers from a KME through a stable interface. ETSI GS QKD 014 is designed to provide that point of interoperability.

\section{THE ETSI GS QKD 014 PROTOCOL}
ETSI GS QKD 014 specifies a REST-based API through which an SAE obtains key material from a KME \cite{etsi014}. The interface uses HTTPS with TLS version 1.2 or higher, mutual authentication between SAE and KME, and JSON request and response bodies. The standard defines three principal methods. First, \texttt{Get status} reports the KME status and key-delivery capabilities for a target slave SAE. Second, \texttt{Get key} allows a master SAE to request one or more fresh keys for a specified slave SAE. Third, \texttt{Get key with key IDs} allows the slave SAE to retrieve keys corresponding to key identifiers previously delivered to the master SAE.

The protocol model is deliberately asymmetric per key. The SAE that initiates \texttt{Get key} is the master SAE for the returned keys. The SAE that later calls \texttt{Get key with key IDs} is the slave SAE for those keys. The returned key container includes an array of key entries, each containing a \texttt{key\_ID} and base64-encoded key material. The standard explicitly leaves the communication of key identifiers between applications outside the API scope. Consequently, any real application must define its own way to carry \texttt{key\_ID} values from the master side to the slave side.

This out-of-scope identifier transport is the main design point of the VPN. For outbound traffic, the local VPN peer acts as the master SAE and requests a 256-bit key through the HTTP route \texttt{POST /api/v1/keys/\{slave\_SAE\_ID\}/enc\_keys}. For inbound traffic, the same VPN process acts as the slave SAE and resolves the sender's key identifier through \texttt{POST /api/v1/keys/\{master\_SAE\_ID\}/dec\_keys}. Thus, each peer performs both ETSI roles, depending on traffic direction.

\chapter{Experiment Design}
The experiment was designed to answer a practical interoperability question: can a simple IP tunnel consume QKD keys through ETSI GS QKD 014 without depending on vendor-specific key files or proprietary control protocols? The experimental artifact consists of two software components and one hardware-facing validation stage.

The first component is the VPN implementation. It is a Linux-only Python prototype that creates a TUN interface, forwards plaintext IP packets over a UDP tunnel, and encrypts each packet using AES-256-GCM. The second component is the \texttt{kms} package, which implements a controlled subset of the ETSI 014 API. The simulator gives the VPN a deterministic interoperability target before the same application is pointed at a real KME. The third stage connects the VPN to the LuxQuanta NOVA QKD backend, using the same ETSI client logic and the same application-level tunnel framing.

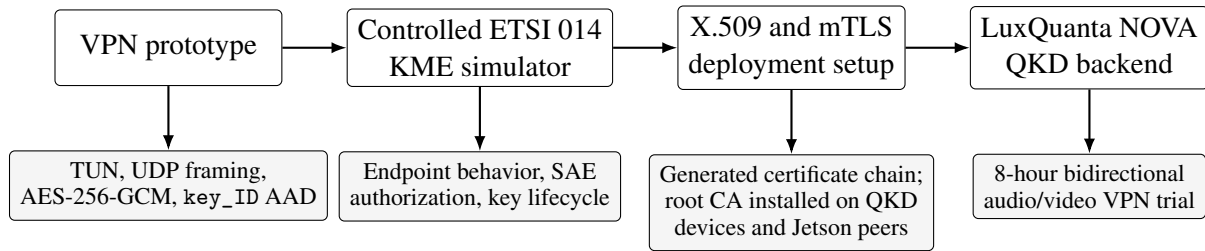
\begin{figure*}[t]
\centering
\begin{tikzpicture}[
    node distance=0.85cm and 0.85cm,
    stage/.style={draw, rounded corners=2pt, align=center, minimum width=3.0cm, minimum height=1.0cm, font=\small},
    artifact/.style={draw, rounded corners=2pt, align=center, minimum width=3.0cm, minimum height=0.8cm, font=\scriptsize, fill=gray!8},
    arrow/.style={-{Latex[length=2mm]}, thick}
]
\node[stage] (vpn) {VPN prototype};
\node[stage, right=of vpn] (sim) {Controlled ETSI 014\\KME simulator};
\node[stage, right=of sim] (pki) {X.509 and mTLS\\deployment setup};
\node[stage, right=of pki] (nova) {LuxQuanta NOVA\\QKD backend};

\node[artifact, below=of vpn] (vpnout) {TUN, UDP framing,\\AES-256-GCM, \texttt{key\_ID} AAD};
\node[artifact, below=of sim] (simout) {Endpoint behavior, SAE\\authorization, key lifecycle};
\node[artifact, below=of pki] (pkiout) {Generated certificate chain;\\root CA installed on QKD\\devices and Jetson peers};
\node[artifact, below=of nova] (novaout) {8-hour bidirectional\\audio/video VPN trial};

\draw[arrow] (vpn) -- (sim);
\draw[arrow] (sim) -- (pki);
\draw[arrow] (pki) -- (nova);
\draw[arrow] (vpn) -- (vpnout);
\draw[arrow] (sim) -- (simout);
\draw[arrow] (pki) -- (pkiout);
\draw[arrow] (nova) -- (novaout);
\end{tikzpicture}
\caption{Methodology overview for the experiment. The software VPN was first analyzed as an ETSI GS QKD 014 consumer, then validated against a controlled KME simulator, and finally exercised against a real QKD backend without changing the application-level tunnel contract.}
\label{fig:methodology}
\end{figure*}

Figure~\ref{fig:methodology} summarizes the experimental method. The design intentionally separates implementation, standards conformance, simulated validation, and hardware-facing interoperability so that failures can be localized either to the VPN application logic, the ETSI API contract, or the deployed QKD backend.

\section{VPN IMPLEMENTATION}
The VPN is a single-process, event-driven user-space tunnel. At startup, it validates the local and peer private IP addresses, requires root privileges for TUN configuration, opens \texttt{/dev/net/tun}, assigns point-to-point addressing, installs a host route to the peer private address, and binds a UDP socket to the configured public endpoint. It then constructs an ETSI 014 client with mutual TLS credentials and performs a \texttt{Get status} request to check that the KME advertises support for the requested 256-bit key size.

The data plane has two symmetric paths. On the outbound path, the VPN reads a packet from the TUN device, checks that it resembles an IPv4 or IPv6 packet, obtains the active transmit key from \texttt{SendState}, constructs a tunnel header, encrypts the packet with AES-GCM, and sends the resulting frame over UDP. \texttt{SendState} holds a single active transmit key lease. If no key has been acquired, or if the configured rotation interval has elapsed, it calls \texttt{Get key} with \texttt{\{"size": 256, "number": 1\}} and replaces the active lease with the returned key and \texttt{key\_ID}.

On the inbound path, the VPN accepts UDP datagrams only from the configured peer public address. It parses the fixed header, extracts the variable-length \texttt{key\_ID}, verifies the protocol version and expected peer sender identifier, checks that the ciphertext length matches the header, rejects repeated sequence numbers for the same \texttt{sender\_id} and \texttt{key\_ID}, retrieves the decryption key through \texttt{ReceiveKeyCache}, and decrypts the frame. Data frames are written into the TUN interface only after successful authentication and after the plaintext again appears to be an IP packet. Keepalive frames are encrypted in the same way as data frames but carry an empty plaintext and are discarded after successful decryption.

The custom frame format is intentionally small. Its fixed header contains a one-byte version, one-byte message type, 32-bit sender identifier, 64-bit sequence number, 16-bit key identifier length, and 32-bit ciphertext length. This fixed header is followed by the UTF-8 encoded ETSI \texttt{key\_ID} and by the AES-GCM ciphertext plus authentication tag. The AES-GCM nonce is the 96-bit concatenation \texttt{sender\_id || seq}. Because the sequence number increases monotonically per process, the construction gives each sender a simple nonce space. The full header, including the \texttt{key\_ID}, is used as AAD. Therefore, the key identifier that selects the decryption key is cryptographically bound to the ciphertext.

The implementation deliberately avoids an additional control channel. ETSI key identifiers are transported in-band as part of every data and keepalive frame. This choice mirrors the ETSI 014 assumption that applications must communicate key identifiers as necessary for their operation. It also simplifies interoperability testing: if the remote KME recognizes the \texttt{key\_ID} and authorizes the receiving SAE, the packet can be decrypted; otherwise, the packet fails closed.

\section{QKD SIMULATION}
The \texttt{kms} package is an implementation of the ETSI-facing behavior needed by the VPN. It exposes routes under \texttt{/api/v1/keys}: \texttt{status}, \texttt{enc\_keys}, and \texttt{dec\_keys}. The simulator supports the mTLS protocol, where the server extracts the SAE identity from the client certificate Common Name and validates it against the configured topology.

The simulator models KME topology through its configuration. Each configured KME owns a set of connected SAEs. A \texttt{Get status} request resolves the calling SAE as the master, resolves the path SAE as the slave, and returns the source and target KME identifiers together with fixed 256-bit key-size capability. A \texttt{Get key} request validates the slave SAE, validates the supported request subset, generates 32 bytes of key material from the operating-system random source, assigns a UUID \texttt{key\_ID}, stores the key with its master and slave SAE ownership metadata, and returns an ETSI key container. A \texttt{Get key with key IDs} request validates that the calling slave SAE and the path master SAE match the ownership recorded when the key was issued. After successful retrieval, the simulator deletes the key from storage, reflecting the ETSI post-condition that delivered keys are removed from the KME key pool.

This simulator was not intended to emulate the quantum optical layer. It emulates the application-facing contract: endpoint structure, JSON field names, key-container encoding, key ownership, SAE authorization, and error behavior for unsupported requests. Its tests exercise the complete issue-and-consume workflow, GET shortcuts for simple requests, rejection of unknown SAEs, rejection of unsupported key sizes and mandatory extensions, and authorization failures when the wrong master or slave attempts to retrieve a key. For the VPN experiment, this provided a controlled reference backend against which the ETSI client and packet-level key identifier transport could be validated before using real QKD infrastructure.

\section{REAL-WORLD TESTING}
The real-world stage preserved the same application architecture and replaced the simulated KME with the QKD backend exposed by the LuxQuanta NOVA platform. The two VPN peers were NVIDIA Jetson Xavier NX Developer Kit systems running Ubuntu 24.04. Jetson A was connected to QKD A, Jetson B was connected to QKD B, and the two QKD peers were connected by the vendor-provided quantum channel and its required classical control/post-processing channel. The VPN treated this QKD layer as opaque and interacted only with the local ETSI GS QKD 014 API exposed beside each QKD peer.

The experiment used two distinct IP planes. The direct classical connection between the Jetsons used subnet \texttt{10.0.0.0/24}, with Jetson A at \texttt{10.0.0.1} and Jetson B at \texttt{10.0.0.2}. The VPN tunnel used subnet \texttt{10.10.0.0/24}, with tunnel addresses \texttt{10.10.0.1} and \texttt{10.10.0.2}. Media applications were configured to use the tunnel addresses, not the direct-link addresses, so the transmitted audio and video traversed the encrypted VPN.

Before running the VPN, an X.509 certificate chain was generated for the deployment. The root certificate was installed on the LuxQuanta NOVA QKD side and on both Jetson peers. Each VPN endpoint used client credentials to authenticate as its local SAE, and the peers verified the KME server certificates through the configured root CA. In this configuration, each VPN endpoint acts as a local SAE connected to its local KME. The VPN authenticates to the KME with client credentials, verifies the KME through the configured CA, calls \texttt{Get status} during startup, obtains fresh encryption keys through \texttt{enc\_keys}, and resolves peer keys through \texttt{dec\_keys}.

\begin{table}[t]
\caption{Representative VPN runtime parameters used in the real QKD experiment. Sensitive deployment-specific identifiers are shown as labels.}
\label{tab:runtime-config}
\centering
\scriptsize
\begin{tabular}{lll}
\hline
Parameter & Jetson A & Jetson B \\
\hline
Direct-link IP & \texttt{10.0.0.1} & \texttt{10.0.0.2} \\
Tunnel IP & \texttt{10.10.0.1} & \texttt{10.10.0.2} \\
UDP listen endpoint & \texttt{10.0.0.1:4444} & \texttt{10.0.0.2:4444} \\
UDP peer endpoint & \texttt{10.0.0.2:4444} & \texttt{10.0.0.1:4444} \\
Local SAE & \texttt{SAE\_A} & \texttt{SAE\_B} \\
Peer SAE & \texttt{SAE\_B} & \texttt{SAE\_A} \\
Sender/peer IDs & \texttt{1}/\texttt{2} & \texttt{2}/\texttt{1} \\
KME endpoint & local QKD A API & local QKD B API \\
Key rotation & \multicolumn{2}{c}{60 s transmit-key lease} \\
Key size & \multicolumn{2}{c}{256-bit AES-GCM key} \\
TUN MTU & \multicolumn{2}{c}{1400 bytes} \\
\hline
\end{tabular}
\end{table}

Each peer was configured with its local tunnel address, peer tunnel address, direct-link UDP endpoint, expected peer UDP endpoint, local and peer SAE identities, sender and peer numeric identifiers, local KME URL, client certificate, client private key, trusted root certificate, 60-second key rotation interval, 256-bit key size, and 1400-byte TUN MTU.

The important experimental constraint is that the VPN does not need to know how NOVA generates, reconciles, or distributes key material between KMEs. Those mechanisms remain below the ETSI boundary. The VPN assumes only the property promised by the KME interface: when one SAE obtains a key and communicates its \texttt{key\_ID}, the authorized peer SAE can retrieve identical key material from its own KME using that identifier. This is precisely the separation of concerns that ETSI 014 is designed to enable.

Testing against the hardware backend therefore focused on interoperability of the application contract. The same packet format used against the simulator was used against the real KME. The same 256-bit key request was issued. The same in-band \texttt{key\_ID} field served as the notification mechanism between master and slave SAEs. The same AES-GCM authentication path verified that both sides had obtained identical key bytes. A successful decrypted packet is consequently evidence that the VPN, the KME API, and the QKD backend agreed on the key identifier, key material, SAE authorization, and traffic direction semantics.

\begin{figure*}[t]
\centering
\begin{tikzpicture}[
    node distance=0.7cm and 1.2cm,
    peer/.style={draw, rounded corners=2pt, align=center, minimum width=2.7cm, minimum height=1.05cm, font=\small, fill=gray!8},
    qkd/.style={draw, rounded corners=2pt, align=center, minimum width=2.7cm, minimum height=1.05cm, font=\small},
    net/.style={align=center, font=\scriptsize},
    arrow/.style={-{Latex[length=2mm]}, thick},
    link/.style={thick}
]
\node[peer] (ja) {Jetson A\\VPN SAE A\\direct: \texttt{10.0.0.1}\\tunnel: \texttt{10.10.0.1}};
\node[peer, right=5.7cm of ja] (jb) {Jetson B\\VPN SAE B\\direct: \texttt{10.0.0.2}\\tunnel: \texttt{10.10.0.2}};
\node[qkd, below=1.4cm of ja] (qa) {QKD A / local KME\\ETSI GS QKD 014\\mTLS API};
\node[qkd, below=1.4cm of jb] (qb) {QKD B / local KME\\ETSI GS QKD 014\\mTLS API};

\draw[link] (ja) -- node[net, above] {direct peer link\\\texttt{10.0.0.0/24}\\UDP tunnel transport} (jb);
\draw[arrow] (ja) -- node[net, left] {\texttt{status}\\\texttt{enc\_keys}\\\texttt{dec\_keys}} (qa);
\draw[arrow] (jb) -- node[net, right] {\texttt{status}\\\texttt{enc\_keys}\\\texttt{dec\_keys}} (qb);
\draw[link] (qa) -- node[net, below] {vendor-provided quantum channel and\\required classical post-processing channel} (qb);

\node[draw, dashed, rounded corners=2pt, fit=(ja)(jb), inner sep=0.25cm, label={[font=\scriptsize]above:Classical VPN data plane}] {};
\node[draw, dashed, rounded corners=2pt, fit=(qa)(qb), inner sep=0.25cm, label={[font=\scriptsize]below:QKD/KME layer}] {};
\end{tikzpicture}
\caption{Topology of the VPN/QKD experiment. Each Jetson Xavier NX Developer Kit ran one VPN peer and connected to its local LuxQuanta NOVA QKD peer. VPN data moved directly between Jetsons over UDP on \texttt{10.0.0.0/24}, while the encrypted inner tunnel used \texttt{10.10.0.0/24}. QKD A and QKD B were connected by the vendor-provided quantum channel and its required classical control/post-processing channel, which the VPN treated as opaque.}
\label{fig:topology}
\end{figure*}
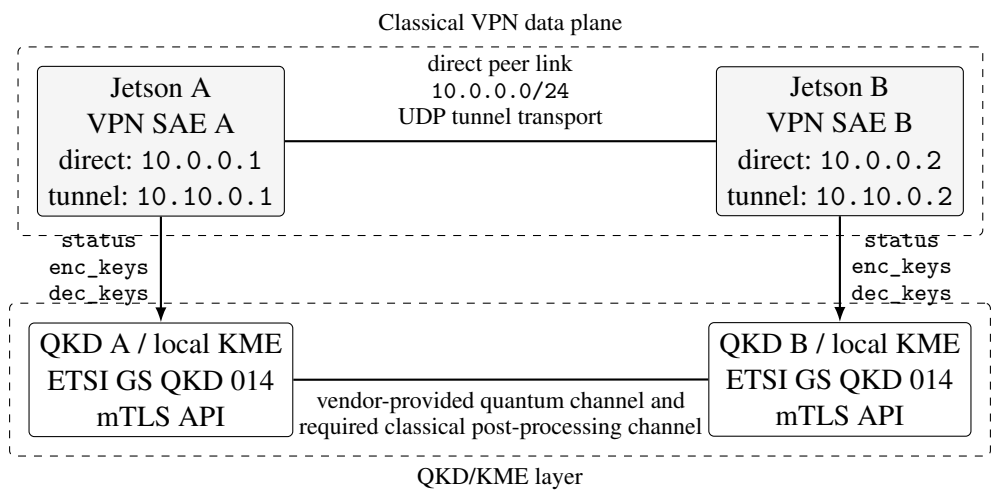

The physical and logical topology used in both simulator and real-backend stages is shown in Fig.~\ref{fig:topology}. The figure emphasizes that the UDP tunnel path and the ETSI key-delivery path are distinct: application traffic flows between VPN peers, while key requests remain local to each peer's KME trust boundary.

\chapter{Results}
The real QKD experiment successfully established an encrypted VPN tunnel between the two Jetson Xavier NX peers using keys delivered by the LuxQuanta NOVA platform through ETSI GS QKD 014. Startup completed on both peers: each VPN process authenticated to its local KME with the deployed certificate chain, verified KME status for the peer SAE, configured its TUN interface, bound its UDP endpoint on the direct peer link, and began exchanging encrypted keepalive and data frames.

The functional validation used real-time audio and video transport to simulate a video-call workload. Each Jetson ran both a sender and a receiver. The sender captured local video from \texttt{/dev/video0} and local audio from ALSA device \texttt{hw:0}, encoded video with H.264 using the \texttt{ultrafast} preset and \texttt{zerolatency} tuning, encoded audio with AAC, and transported the result as MPEG-TS over UDP. The receiver used \texttt{ffplay} with low-buffering options to play the incoming stream. The application-level UDP destinations were the remote VPN tunnel addresses: Jetson A sent to \texttt{10.10.0.2:1234}, and Jetson B sent to \texttt{10.10.0.1:1234}. Both receivers listened locally on UDP port \texttt{1234}.

\begin{lstlisting}[caption={Media transport commands used on both Jetson peers. Jetson A set DEST\_IP=10.10.0.2; Jetson B set DEST\_IP=10.10.0.1.},label={lst:ffmpeg-commands}]
# Sender, with DEST_IP set to the remote tunnel IP
ffmpeg -f v4l2 -video_size 800x600 -i /dev/video0 \
  -f alsa -i hw:0 -c:v libx264 -preset ultrafast \
  -tune zerolatency -c:a aac -f mpegts \
  udp://DEST_IP:1234?pkt_size=1316

# Receiver
ffplay -fflags nobuffer -flags low_delay -probesize 32 \
  -analyzeduration 0 -i udp://@:1234
\end{lstlisting}

The simultaneous bidirectional media test ran continuously for eight hours. During this period, both peers transmitted and received real-time audio and video through the VPN. The observed video quality was good for the feasibility objective, and playback remained usable in real time. Because the media destinations were the \texttt{10.10.0.0/24} tunnel addresses, all exchanged media packets entered the TUN interfaces and were encrypted by the VPN before crossing the direct \texttt{10.0.0.0/24} peer link.

The default 60-second transmit-key rotation interval was used throughout the test. Consequently, each direction requested approximately one fresh 256-bit AES-GCM key per minute from its local KME while traffic was active, and the opposite peer resolved the corresponding \texttt{key\_ID} through \texttt{dec\_keys}. No evidence of key-delivery saturation was observed during this workload: \texttt{enc\_keys} requests continued to succeed, receiving peers continued to resolve peer \texttt{key\_ID} values, and no media interruption was attributed to KME key exhaustion.

\chapter{Discussion}
The experiment confirms an important architectural point: ETSI GS QKD 014 is sufficient as an application-facing key delivery protocol for a VPN-like consumer, provided that the application supplies its own identifier transport and cryptographic binding. The VPN's in-band \texttt{key\_ID} field fills the notification gap left intentionally outside the standard, while AES-GCM AAD prevents an attacker from modifying that field without detection.

The eight-hour media test gives practical evidence for this integration pattern. The media workload was intentionally conventional: standard Linux capture devices, standard \texttt{ffmpeg}/\texttt{ffplay} tooling, UDP transport, and a low-latency encoding profile. The QKD-specific behavior was confined to the VPN and KME boundary. From the media application's point of view, the remote peer was simply reachable through \texttt{10.10.0.0/24}; from the VPN's point of view, fresh symmetric keys and remote key resolution were available through ETSI 014; and from the QKD device's point of view, the Jetsons were authenticated SAEs requesting and consuming keys.

The result also clarifies what was, and was not, demonstrated. The prototype used QKD-delivered symmetric keys as periodically rotated AES-256-GCM keys. It did not implement a one-time-pad design and did not use a distinct QKD key for every media packet. With the 60-second lease, each key protected multiple tunnel frames in one direction until the next rotation. This is appropriate for validating ETSI key delivery in a VPN prototype, but a production design would need a formal key-use policy based on volume, time, failure handling, and nonce persistence.

The KME simulator remains an important component even though the final feasibility test used real LuxQuanta NOVA hardware. The simulator provides a controlled development target for QKD-based applications: developers can implement ETSI endpoint handling, SAE identity policy, key identifier transport, and error handling locally, then swap the backend for a real QKD device without changing the application-facing contract. In this experiment, that staged approach reduced ambiguity before hardware testing because the VPN's use of \texttt{status}, \texttt{enc\_keys}, \texttt{dec\_keys}, and in-band \texttt{key\_ID} transport had already been exercised against a known reference environment.

Several limitations remain. The tunnel has no peer handshake, no negotiated parameter set, no recovery protocol, no congestion control, and no NAT traversal. Its replay protection records seen sequence numbers in a time-bounded set, which is appropriate for a prototype but not optimized for high-rate or long-lived production traffic. The receive-side cache is necessary because the simulator consumes keys after \texttt{dec\_keys}; however, cache lifetime and key-use limits would require more precise policy in a production design. Finally, the nonce construction depends on the uniqueness of \texttt{sender\_id} and monotonic sequence numbers across the lifetime of a key. While this prototype was enough for this experiment, a production implementation would need persistent sender state or an explicit session mechanism to protect against nonce reuse after restart.

Despite these limitations, the experiment is valuable because it isolates the essential interoperability mechanism. The VPN does not embed QKD-specific optical assumptions. It consumes ETSI key containers, transports \texttt{key\_ID} values, and uses the retrieved bytes as ordinary symmetric keys. This is the practical form in which many classical applications are likely to integrate QKD systems.

\chapter{Conclusion}
This experiment implemented and evaluated a prototype VPN that consumes QKD-derived keys through the ETSI GS QKD 014 API. The work connects three layers that are often discussed separately: a classical IP tunnel, an authenticated encryption data plane, and a standardized QKD key delivery interface. The local KMS simulator enabled controlled validation of ETSI endpoint behavior, SAE authorization, key ownership, and one-time retrieval semantics. The subsequent LuxQuanta NOVA test exercised the same application contract against real QKD infrastructure.

The resulting design demonstrates a coherent integration pattern. A VPN peer requests transmit keys as an ETSI master SAE, embeds the returned \texttt{key\_ID} in authenticated tunnel frames, and retrieves peer keys as an ETSI slave SAE. This pattern respects the ETSI boundary: QKD generation and inter-KME synchronization remain the responsibility of the QKD system, while the application handles packet protection and key identifier notification. Future work should turn this validated prototype into a measured experimental study and then into a more robust protocol with explicit sessions, negotiated capabilities, persistent nonce safety, and production-grade operational controls.



\bookmarksetup{startatroot}%

\postextual

\bibliography{references}




\begin{apendicesenv}


\end{apendicesenv}


\begin{anexosenv}



\end{anexosenv}

\end{document}

%% file: tex/acronimos.tex

\newacronym{OTP}{OTP}{Chave de Uso Único (do inglês, \textit{One-Time Pad})}
\newacronym{AES}{AES}{Padrão de Criptografia Avançada (do inglês, \textit{Advanced Encryption Standard})}
\newacronym{ETSI}{ETSI}{Instituto Europeu de Normas de Telecomunicações (do inglês, \textit{European Telecommunications Standards Institute})}
\newacronym{PSK}{PSK}{Chave Pré-Compartilhada (do inglês, \textit{Pre-Shared Keys})}
\newacronym{PQC}{PQC}{Criptografia Pós-Quântica (do inglês, \textit{Post-Quantum Cryptography})}
\newacronym{DM-CVQKD}{DM-CVQKD}{Distribuição Quântica de Chaves com Modulação Discreta em Variáveis Contínua (do inglês, \textit{Discrete-Modulated Continuous-Variable Quantum Key Distribution})}
\newacronym{MitM}{MitM}{Homem no Meio (do inglês, \textit{Man-in-the-Middle})}
\newacronym{TLS}{TLS}{\textit{Transport Layer Security}}
\newacronym{SSL}{SSL}{\textit{Secure Sockets Layer}}
\newacronym{IETF}{IETF}{Força Tarefa de Engenharia de Internet (do inglês, \textit{(Internet Engineering Task Force})}
\newacronym{TCP}{TCP}{Protocolo de Controle de Transmissão (do inglês, \textit{Transmission Control Protocol})}
\newacronym{IPsec}{IPsec}{\textit{Internet Protocol Security}}
\newacronym{VPN}{VPN}{Rede Privada Virtual (do inglês, \textit{Virtual Private Network)}}
\newacronym{IKE}{IKE}{\textit{Internet Key Exchange}}
\newacronym{SA}{SA}{Associação de Segurança (do inglês, \textit{Security Associations})}
\newacronym{AH}{AH}{\textit{Authentication Header}}
\newacronym{ESP}{ESP}{\textit{Encapsulating Security Payload}}
\newacronym{ITU}{ITU}{União Internacional de Telecomunicações (do inglês, \textit{International Telecommunication Union})}
\newacronym{SAE}{SAE}{\textit{Secure Application Entity}}
\newacronym{KME}{KME}{\textit{Key Management Entity}}
\newacronym{HTTPS}{HTTPS}{Protocolo de Transferência de Hipertexto Seguro (do inglês, \textit{Hypertext Transfer Protocol Secure})}
\newacronym{JSON}{JSON}{\textit{JavaScript Object Notation}}
\newacronym{KM}{KM}{\textit{Key Manager}}
\newacronym{QoS}{QoS}{Qualidade de Serviço (do inglês, \textit{Quality of Service})}
\newacronym{KMS}{KMS}{Sistema de Gerenciamento de Chaves (do inglês, \textit{Key Management System})}
\newacronym{HSM}{HSM}{Módulo de Segurança de Hardware (do inglês, \textit{Hardware Security Module})}
\newacronym{SDN}{SDN}{Redes Definidas por Software (do inglês, \textit{Software-Defined Networking})}
\newacronym{PUF}{PUF}{\textit{Physical Unclonable Function}}
\newacronym{SSH}{SSH}{\textit{Secure Shell}}
\newacronym{MQTT}{MQTT}{\textit{Message Queuing Telemetry Transport}}
\newacronym{IoT}{IoT}{Internet das Coisas (do inglês, \textit{Internet of Things})}
\newacronym{SMTP}{SMTP}{Protocolo de Transferência de Correio Simples (do inglês, \textit{Simple Mail Transfer Protocol})}
\newacronym{MLS}{MLS}{Segurança da camada de mensagens (do inglês, \textit{Messaging Layer Security})}
\newacronym{MACsec}{MACsec}{\textit{Media Access Control Security}}
\newacronym{MKA}{MKA}{\textit{MACsec Key Agreement}}
\newacronym{TCP/IP}{TCP/IP}{\textit{Transmission Control Protocol/Internet Protocol}}
\newacronym{UDP}{UDP}{\textit{User Datagram Protocol}}
\newacronym{IP}{IP}{\textit{Internet Protocol}}
\newacronym{DNS}{DNS}{Sistema de Nomes de Domínio (do inglês, \textit{Domain Name System})}
\newacronym{MAC}{MAC}{\textit{Media Access Control}}

\newacronym{RAM}{RAM}{Memória de Acesso Aleatório (do inglês Random Access Memory)}
\newacronym{ASIC}{ASIC}{Circuito Integrado para Aplicação Específica (do inglês \textit{Application-specific Integrated Circuit})}
\newacronym{ADC}{ADC}{Conversores Analógico-Digital (do inglês \textit{Analog-to-Digital Converter})}
\newacronym{AWG}{AWG}{Gerador de Formas de Onda Arbitrárias (do inglês, \textit{Arbitrary Waveform Generator})}
\newacronym{AWGN}{AWGN}{Ruído Branco Gaussiano Aditivo (do inglês, \textit{Additive White Gaussian Noise})}
\newacronym{BB84}{BB84}{QKD desenvolvido por Charles Bennett e Gilles Brassard em 1984}
\newacronym{BER}{BER}{Taxa de Erro de Bit (do inglês, \textit{Bit Error Rate})}
\newacronym{BHD}{BHD}{Fotodetectores Balanceados (do inglês, \textit{Balanced Homodyne Detection})}
\newacronym{BP}{BP}{Propagação de Crenças (do inglês, \textit{Belief Propagation})}
\newacronym{BPSK}{BPSK}{Chaveamento por Deslocamento de Fase Binária (do inglês, \textit{Binary Phase Shift Keying})}
\newacronym{BS}{BS}{Divisor de Feixe (do inglês, \textit{Beam Splitter})}
\newacronym{CAZAC}{CAZAC}{Amplitude Constante e Autocorrelação Zero (do inglês, \textit{Constant Amplitude Zero Auto-Corelation})}
\newacronym{CCDM}{CCDM}{Casamento de Distribuição por Composição Constante (do inglês, \textit{Constant Composition Distribution Matching})}
\newacronym{CD}{CD}{Dispersão Cromática (do inglês, \textit{Chromatic Dispersion})}
\newacronym{CMRR}{CMRR}{Relação de Rejeição de Modo Comum (do inglês, \textit{Common Mode Rejection Ratio})}
\newacronym{CSPRNGs}{CSPRNGs}{Geradores de Números Pseudoaleatórios Criptograficamente Seguros (do inglês, \textit{Cryptographically Secure PRNGs})}
\newacronym{CV-QKD}{CV-QKD}{Distribuição Quântica de Chaves com Variáveis Contínuas (do inglês, \textit{Continuous Variable Quantum Key Distribution})}
\newacronym{CPE}{CPE}{Estimação de Fase da Portadora (do inglês, \textit{Carrier Phase Estimation})}
\newacronym{CPU}{CPU}{Unidade Central de Processamento (do inglês, \textit{Central Unit Processor})}
\newacronym{DFB}{DFB}{Lasers de Realimentação Distribuída (do inglês, \textit{Distributed-Feedback Lasers})}
\newacronym{DM}{DM}{Casador de Distribuição (do inglês, \textit{Distribution Matcher})}
\newacronym{DPM}{DPM}{Modulador de Dupla Polarização (do inglês, \textit{Dual-Polarization Modulator})}
\newacronym{DR}{DR}{Reconciliação Direta (do inglês, \textit{Direct Reconciliation})}
\newacronym{DAC}{DAC}{Conversores Digital-Analógico (do inglês, \textit{Digital-to-Analog Converter})}
\newacronym{DV-QKD}{DV-QKD}{QKD com Variáveis Discretas (do inglês, \textit{Discrete Variable Quantum Key Distribution})}
\newacronym{DSP}{DSP}{Processamento Digital de Sinal (do inglês, \textit{Digital Signal Processing})}
\newacronym{ECC}{ECC}{Códigos de Correção de Erro (do inglês, \textit{Error Correction Code})}
\newacronym{ECLs}{ECLs}{Lasers de Cavidade Externa (do inglês, \textit{External-Cavity Lasers})}
\newacronym{EVOA}{EVOA}{Atenuador Óptico Variável Eletrônico (do inglês, \textit{Electronic Variable Optical Attenuator})}
\newacronym{FER}{FER}{Taxa de Erro de Frame (do inglês, \textit{Frame Error Rate})}
\newacronym{FIR}{FIR}{Resposta ao Impulso Finito (do inglês, \textit{Finite Impulse Response})}
\newacronym{FPGA}{FPGA}{Arranjo de Porta Programável em Campo (do inglês, \textit{Field Programmable Gate Array})}
\newacronym{GMCS}{GMCS}{Modulação Gaussiana de Estados Coerentes (do inglês, \textit{Gaussian Modulation of Coherent States})}
\newacronym{GPGPU}{GPGPU}{Unidade de processamento Gráfico de Proposito Geral (do inglês, \textit{General Purpose Graphical Processor Unit})}
\newacronym{HDL}{HDL}{Linguagem de Descrição de Hardware (do inglês, \textit{Hardware Description Language})}
\newacronym{HPC}{HPC}{Computador de Alto Desempenho (do inglês, \textit{High Performance Computer})}
\newacronym{IIR}{IIR}{Resposta ao Impulso Infinito (do inglês, \textit{Infinite Impulse Response})}
\newacronym{IQ}{IQ}{Em Fase I e Quadratura Q (do inglês, \textit{In-phase I and Quadrature Q})}
\newacronym{IQM}{IQM}{Modulador em Fase e Quadratura (do inglês, \textit{In-phase and quadrature Modulator})}
\newacronym{IR}{IR}{Reconciliação de Informação (do inglês, \textit{Information Reconciliation})}
\newacronym{IRforCV-QKD}{IRforCV-QKD}{Biblioteca de Reconciliação de Informações para Sistemas CV-QKD (do inglês, \textit{Information Reconciliation Library for CV-QKD Systems})}
\newacronym{ISI}{ISI}{Interferências Entre Símbolos (do inglês, \textit{Intersymbol Interference})}
\newacronym{LDPC}{LDPC}{Verificação de Paridade de Baixa Densidade (do inglês, \textit{Low Density Parity Check})}
\newacronym{LFSR}{LFSR}{Registradores de Deslocamento de Realimentação Linear (do inglês, \textit{Linear Feedback Shift Registers})}
\newacronym{LLO}{LLO}{Oscilador Local localmente gerado (do inglês, \textit{Locally generated Local Oscillator})}
\newacronym{LLR}{LLR}{Razão de Verossimilhança Logarítmica (do inglês, \textit{Log Likelihood Ratio})}
\newacronym{LLR-BP}{LLR-BP}{Razão de Verossimilhança Logarítmica em Propagação de Crenças (do inglês, \textit{Log-Likelihood Ratio BP})}
\newacronym{LMS}{LMS}{Mínimos Quadrados Médios (do inglês, \textit{Least Mean Squares})}
\newacronym{LO}{LO}{Oscilador Local (do inglês, \textit{Local Oscilator})}
\newacronym{M-APSK}{M-APSK}{Chaveamento por Deslocamento de Fase e Amplitude M-ária (do inglês, \textit{M-ary Amplitude and Phase Shift Keying})}
\newacronym{MLE}{MLE}{Estimador de Máxima Verossimilhança (do inglês, \textit{Maximum Likelihood Estimator})}
\newacronym{ML}{ML}{Aprendizado de Máquina (do inglês, \textit{Machine Learning})}
\newacronym{MLPs}{MLPs}{Perceptron Multicamadas (do inglês, \textit{Multi-layer Perceptron})}
\newacronym{M-PAM}{M-PAM}{Modulação por Amplitude de Pulso M-ária (do inglês, \textit{M-ary Pulse Amplitude Modulation})}
\newacronym{MSE}{MSE}{Erro Quadrático Médio (do inglês, \textit{Mean Squared Error})}
\newacronym{MSD}{MSD}{Decodificador Multi-Estágio (do inglês, \textit{Multi-Stage Decoder})}
\newacronym{MZM}{MZM}{Modulador Mach-Zehnder (do inglês, \textit{Mach-Zehnder Modulator})}

\newacronym{PA}{PA}{Amplificação de Privacidade (do inglês, \textit{Privacy Amplification})}
\newacronym{PBC}{PBC}{Combinador de Feixe de Polarização (do inglês, \textit{Polarization Beam Combiner})}
\newacronym{PBS}{PBS}{Divisor de Feixe de Polarização (do inglês, \textit{Polarizing Beam Splitter})}
\newacronym{PC}{PC}{Controlador de Polarização (do inglês, \textit{Polarization Controller})}
\newacronym{PCDM}{PCDM}{Casamento de Distribuição por Composição Constante Paralelo (do inglês, \textit{Parallel CCDM})}
\newacronym{PCM}{PCM}{Matriz de Verificação de Paridade (do inglês, \textit{Parity-Check Matrix})}
\newacronym{PM}{P\&M}{Preparar e Medir (do inglês, \textit{Prepare and Measure})}
\newacronym{EB}{EB}{Baseado em Emaranhamento (do inglês, \textit{Entanglement Based})}
\newacronym{PRNGs}{PRNGs}{Geradores de Números Pseudoaleatórios (do inglês, \textit{Pseudo-Random Number Generators})}
\newacronym{PSA}{PSA}{Amplificador Sensível à Fase (do inglês, \textit{Phase-Sensitive Amplifier})}
\newacronym{QAM}{QAM}{Modulação de Amplitude em Quadratura (do inglês, \textit{Quadrature Amplitude Modulation})}
\newacronym{QKD}{QKD}{Distribuição Quântica de Chaves (do inglês, \textit{Quantum Key Distribution})}
\newacronym{QOSST}{QOSST}{Software Aberto para Transmissões Quânticas Seguras (do inglês, \textit{Quantum Open Software for Secure Transmissions})}
\newacronym{QPSK}{QPSK}{Modulação por Deslocamento de Fase em Quadratura (do inglês, \textit{Quadrature Phase Shift Keying})}
\newacronym{QRNG}{QRNG}{Gerador Quântico de Números Aleatórios (do inglês, \textit{Quantum Random Number Generation})}
\newacronym{RC}{RC}{Cosseno Levantado (do inglês, \textit{Raised Cosine})}
\newacronym{RIN}{RIN}{Ruído de Intensidade Relativa (do inglês, \textit{Relative Intensity Noise})}
\newacronym{RR}{RR}{Reconciliação Reversa (do inglês, \textit{Reverse Reconciliation})}
\newacronym{RRC}{RRC}{Raiz do Cosseno Levantado (do inglês, \textit{Root-Raised-Cosine})}
\newacronym{RTL}{RTL}{Nível de Transferência entre Registradores (do inglês, \textit{Register Transfer Level})}
\newacronym{SER}{SER}{Taxa de Erro de Símbolo (do inglês, \textit{Symbol Error Rate})}
\newacronym{SKR}{SKR}{Taxa de Chave Secreta (do inglês, \textit{Secure Key Rate})}
\newacronym{SLMs}{SLMs}{Lasers de Modo Longitudinal Único (do inglês, \textit{Single Longitudinal-Mode Lasers})}
\newacronym{SNR}{SNR}{Relação Sinal-Ruído (do inglês, \textit{Signal-to-Noise Ratio})}
\newacronym{SNU}{SNU}{Unidade de Shot-Noise (do inglês, \textit{Shot-Noise Unit})}
\newacronym{SPS}{SPS}{Amostras Por Símbolo (do inglês, \textit{Samples-Per-Symbol})}
\newacronym{TIA}{TIA}{Amplificadores de Transimpedância (do inglês, \textit{Transimpedance Amplifiers})}
\newacronym{VOA}{VOA}{Atenuador Óptico Variável (do inglês, \textit{Variable Optical Attenuator})}
\newacronym{ZC}{ZC}{Sequência Zadoff-Chu}
\newacronym{ZF}{ZF}{Força Zero (do inglês, \textit{Zero-Forcing})}
\newacronym{SOC}{SoC}{Sistema em um Chip (do inglês, \textit{System on Chip})}
\newacronym{GPU}{GPU}{Unidade de Processamento Gráfico (do inglês, \textit{Graphics Processing Unit})}
\newacronym{PID}{P,D\&I}{Pesquisa, Desenvolvimento e Inovação}
\newacronym{CLB}{CLB}{Blocos Lógicos Configuráveis (do inglês, \textit{Configurable Logic Blocks})}
\newacronym{LE}{LE}{Elementos Lógicos (do inglês, \textit{Logic Elements})}
\newacronym{LUT}{LUT}{tabela de consulta (do inglês, \textit{Look Up Table})}
\newacronym{HLS}{HLS}{síntese de alto nível (do inglês, \textit{High-Level Synthesis})}
\newacronym{QC}{QC}{Quase Cíclico}
\newacronym{MET}{MET}{Multi Edge}
\newacronym{GSPS}{GSPS}{Giga Amostras por Segundo (do inglês, \textit{Gigasamples per Second})}
\newacronym{bps}{bps}{bits por segundo (do inglês, \textit{bits per second})}
\newacronym{GF(2)}{GF(2)}{Campo de Galois ou corpo finito 2 (do inglês, \textit{Galois field})}
\newacronym{CI}{CI}{Circuito Integrado}
\newacronym{DSE}{DSE}{Exploração do Espaço de Projeto (do inglês, \textit{Design Space Exploration})}

%% file: tex/glossario.tex
\newglossaryentry{latex}
{
        name=latex,
        description={Is a mark up language specially suited for 
scientific documents}
}

\newglossaryentry{maths}
{
        name=mathematics,
        description={Mathematics is what mathematicians do}
}

\newglossaryentry{formula}
{
        name=formula,
        description={A mathematical expression}
}